\documentstyle[aps, preprint]{revtex}
\tightenlines

\def\be{\begin{equation}}
\def\ee{\end{equation}}
\def\ba{\begin{eqnarray}}
\def\ea{\end{eqnarray}}

\def\k{\mbox{\bf k}}
\def\p{\mbox{\bf p}}

\begin{document}
\input epsf
\draft
\preprint{PURD-TH-00-05, hep-ph/0009093}
\title{Large Metric Perturbations from Rescattering}

\author{F. Finelli and S. Khlebnikov}

\address{Department of Physics, Purdue University,
West Lafayette,
IN 47907, USA}

\maketitle

\begin{abstract}
We study numerically evolution of metric perturbations during
reheating in a model with two fields and a strong parametric resonance.
Our calculation is fully nonlinear and includes gravity but is restricted
to spherical symmetry. In this model, super-Hubble metric perturbations can grow
during reheating only due to effects nonlinear in fluctuations of the fields. 
We find that they indeed grow and, soon after the growth begins, dominate variances 
of the metric functions. Thus, the metric functions become smooth but varying 
significantly over large scales. Their profiles at late times are interpreted 
as signalling a gravitational instability and formation of a black hole.

\end{abstract}


%\newpage

\narrowtext
Inflationary cosmology (reviewed in books \cite{books}) provides a very attractive 
explanation of the
origin of the large-scale structure of the universe \cite{structure}. If the
only important 
interaction of the scalar field (the inflaton) is gravity, metric perturbations 
at super-Hubble
scales can grow during inflation and during the transition to 
radiation-dominated
(RD) era, but not during the RD era itself. Thus, they are preserved until
the time when a gravitational instability is ready to convert them into 
the observed structure. 

The question remains, however, whether that scenario can
be changed by additional interactions of the inflaton (a self-interaction
or interactions with other fields).
Specifically, right after the inflation, at {\em reheating},
the inflaton still has 
a large homogeneous component, and it is easy to imagine that
fluctuations
of various fields (and essentially of any scale) produced at that time
will kick particles out of the homogeneous modes, into modes with
large, super-Hubble wavelengths.

A natural arena to address the above question is inflationary models in which 
reheating starts out explosively, via a parametric resonance \cite{param}.
In rigid spacetimes, the growth of large-wavelength modes due to rescattering
\cite{KT1}, and the resulting formation of smooth (turbulent) power spectra,
are clearly seen in a variety of inflationary models
\cite{KT1,variety}. To extend these analyses to super-Hubble scales,
it is necessary to include gravity, i.e. perturbations of the metric.

The influence of parametric resonance on the evolution of metric perturbations
at first order in field fluctuations was addressed in 
\cite{BKM,FB1,BV,FB2,ZBS}.
While it is possible to identify models in which the growth of
super-Hubble metric perturbations occurs at that order \cite{BV,FB2}, 
in many popular 
models such growth (if any) can only 
appear at the second order (e.g. through the scattering process 
described above). Such second-order effects are quite generic.
Fortunately, models with a parametric resonance are amenable to a fully
nonlinear numerical study, based on the classical approximation \cite{KT1}.

Here we present results of such a study, including gravity but restricted to
spherical symmetry, for the model of two interacting field with the potential
\be
V(\phi, \chi) = g^2 \phi^2 \chi^2 / 2 + \lambda \phi^4 / 4 \; ,
\label{pot}
\ee
where $\phi$ is the inflaton, and $\chi$ is a field that initially has
no homogeneous component. Both fields are minimally coupled to gravity.

We choose one the ``magic numbers'' for $g^2/\lambda$, 
for which the resonance is especially strong \cite{GKLS}: $g^2/\lambda =
8$. With this choice, the resonance
peak is at $k=0$, so super-Hubble fluctuations of $\chi$ will start 
growing right away.
Because $\chi$ initially has no homogeneous component, that growth will
induce fluctuations of the metric that are still of the second
order in $\delta \chi$.
We do not believe the choice of $g^2 /\lambda$ to be of great importance
(although a weaker resonance or a smaller initial $\delta \chi$ 
will delay the onset of nonlinearity in field fluctuations); we tried other values, 
with very similar results \cite{long}. Note that the scattering mechanism 
described above will work even in 
the absence of any preexisting fluctuations of $\chi$ at super-Hubble scales.

We find that (i) super-Hubble perturbations of metric indeed grow, and 
(ii) profiles of the metric functions eventually undergo a significant change,
which we interpret as development of gravitational instability and formation of
a black hole.

The first of these results is in agreement with a result obtained
by Easther and Parry for the $\lambda\phi^4$ model in planar geometry
\cite{EP2}, but the second one is not. We do not think that the difference
is due to the different
choice of a model, but are inclined to attribute it either to the difference
in geometry or to the presence of an additional radiation field in their 
calculation.

When we speak about terms of the second order in $\delta\chi$, we 
envision fluctuations of the energy density of the form
\be
\delta E_{\k} \propto 
\int d^3 p F(\p, \k) \delta \chi_{\p} \delta \chi_{\p + \k} \; ,
\label{dE}
\ee
with nonzero $\k$ ($F$ is some function; a corresponding term with
$\k=0$ does not contribute to $\delta E$, but only to the average energy).
Terms like (\ref{dE}) appear both directly from $(\delta\chi)^2$ terms in 
the energy, and from terms proportional to $\delta\phi$, after a sizable
$\delta\phi$ is generated by rescattering. 

We caution, however, against relying on the size of $\delta E$ as the criterion
for formation of black holes, as seems to be the trend in recent
literature \cite{GMBT}.
One needs to solve explicitly for the metric functions.
In Newtonian gravity, the gravitational potential will typically be smoother
that the mass distribution. Similarly, in our simulations (using general
relativity), we find 
that while the profiles of the fields $\phi$ and $\chi$ 
(and of the spatial curvature), at the
time when a black hole forms, are fairly rugged, those of the metric functions
are exceptionally smooth. We stress that at that time the power spectra of the 
fields and of the metric functions are determined by rescattering, rather than
by their initial spectra or the original shape of the resonance. Thus, rescattering
becomes a crucial effect. 

We use a spherically symmetric metric in isotropic coordinates: 
\be
ds^2 = - N^2(t,r) dt^2 + \psi^4(t, r) \left[ dr^2 + r^2 d\Omega^2 \right]
\; .
\label{met}
\ee
Equations of motion are written in the Hamiltonian form, the Hamiltonian pairs
being $(\phi, \pi_\phi)$, $(\chi, \pi_\chi)$, and $(\psi, K)$, where $K$
is the trace of
the extrinsic curvature: $K = -6\dot{\psi}/ N\psi$. 
The lapse $N$ does not have a
pair; it satisfies an ordinary differential equation that involves only spatial
derivatives. 
The energy and momentum of the scalar fields are
\begin{eqnarray}
E & = & (\pi_\phi^2 + \pi_\chi^2) / 2\psi^{12} 
+ ({\phi'}^2 + {\chi'}^2) / 2\psi^4 + V(\phi, \chi) \; , \label{E} \\
P & = & (\pi_\phi \phi' + \pi_\chi \chi') / \psi^6 \; , \label{P} 
\end{eqnarray}
and the energy and momentum constraints are
\begin{eqnarray}
C_E & \equiv & R + 8 \nabla^2\psi / \psi^5 = 0 \; ,
\label{ec} \\
C_M & \equiv & P - 2 K' / 3 \kappa^2 = 0 \; , \label{mc}
\end{eqnarray}
where $R$ is the curvature of the spatial sections:
\be
R/2 = \kappa^2 E - K^2 / 3 \; ,
\label{R}
\ee
and $\kappa^2 = 8\pi G$.

We take the equation for $K$ in the form
\be
\frac{\dot K}{N} = \frac{K^2}{2} -6 \frac{\psi'}{\psi^5}
\left( \frac{\psi'}{\psi} + \frac{1}{r} \right) - 3 \frac{N'}{N \psi^4}
\left( \frac{1}{r} + 2 \frac{\psi'}{\psi} \right) 
+ \frac{3}{2} \kappa^2 \left( E - 2 V  \right) - \frac{3}{4} w C_E \; .
\label{dotK}
\ee
Adding the constraint function $C_E$ with some coefficient $w > 0$ supplies
the equation with a second spatial derivative of $\psi$ 
and, for a range of $w$, prevents
a short-scale numerical instability. We present results for $w=3$; results
for $w=1$ are similar. All other equations are standard, and we do not 
write them here.
 
It is convenient to work
in rescaled variables, in which $\phi$ and $\chi$ are measured in units of 
$\phi(0)$, the field at the end of inflation, while time and distance
are measured in units of $1/\sqrt{\lambda} \phi(0)$. Comoving momenta $k$ are
in units of $\sqrt{\lambda} \phi(0)$. The evolution equations retain 
their form, with the following replacements: 
$\lambda\to 1$, $g^2\to g^2/\lambda$, 
and $\kappa^2 \to \kappa^2 \phi^2(0) = 3.08$. 
The numerical value for $\phi(0)$ is adopted from the previous work
\cite{KT1}.

In the classical approximation, fields are regarded
as classical entities with random initial conditions. Their Fourier components
in the initial state are Gaussian and distributed according to certain power
spectra. We choose these initial spectra as follows.

Amplification of quantum fluctuations of the metric during inflation results 
in an approximately scale-invariant spectrum for $\psi$: 
$\langle\langle |\psi_{\k}|^2 \rangle\rangle \propto 1/k^3$; 
double brackets denote
averaging over the quantum state. Then, from the energy
constraint, $\langle\langle |\pi_{\phi\k}|^2 \rangle\rangle \propto k$
at large $k$. 
We choose to cut off the spectra at large $k$ to remove the corresponding
large contribution to energy, so it does not cause spurious gravitational
effects at early times. 
Thus, we use for $\psi$ an initial power spectrum of the form 
$k^{-3}\exp(-k^2/\mu^2)$, with $\mu^2 = 10$. This cutoff is imposed only
at the initial instant, and the results at large times are insensitive to
it. The overall magnitude of the $\psi$ spectrum is chosen to get 
a realistic size of fluctuations on super-Hubble scales, 
$\delta\psi / \psi \sim 10^{-5}$.

Initial spectra of $\phi$, $\chi$, and $K$ are then obtained as follows.
Fluctuations of $\chi$ are placed in adiabatic vacuum, with a cutoff factor
as above. This leads to an initial $\delta\chi \sim 3\times 10^{-8} M_{\rm Pl}$.
Fluctuations of $\phi$ are determined from the constraints. We either
solve the constraints to the first order, or use a trivial exact solution:
$\delta\phi = 0$, with $\delta K$ determined from (\ref{mc}), and
$\delta \pi_{\phi}$ from (\ref{ec}).
Results for these two types of initial conditions are quite similar; here
we present those for the exact solution.
 
The Hamiltonian structure of the equations suggests that it is suitable to use 
a leapfrog algorithm,
in which the coordinates are viewed as defined at full time steps, and
the momenta at half-steps. The lapse $N$ is updated from its 
differential equation
at full steps and is extrapolated to half-steps.
The equation for $N$ does not fix its overall normalization, so $N$ can 
be arbitrarily scaled at each spatial section to define a 
convenient time variable. 
We use conformal time, obtained by setting $N$ at the outer 
end of the grid equal
to the scale factor (defined below).
We use the boundary condition $f_n = f_{n - 1}$ at the outer end;
$f$ is any field or metric function.

At times shown in the figures, $\psi_n^2$, the square of $\psi$ at the end of 
the grid, grows steadily with time and even at the latest times
deviates relatively little from a linear $t$ dependence corresponding to 
an RD universe. Motivated by that, we adopted the 
following definitions of the scale factor $a(t)$ and the comoving
Hubble parameter $H(t)$: $a = \psi_n^2$, $H= - a K_n / 3$.

We output two types of quantities as functions of the conformal time. 
Some are built from ordinary lattice averages, such as
\be
\langle f \rangle = S_2^{-1} \sum_{i = 1}^{n-1} i^2 f_i \; ,
\label{ave}
\ee
where $S_2 = (2n^3 - 3n^2 + n) / 6$.
These include ordinary root mean square (r.m.s.) deviations---square roots of
the variances $\langle (\delta f)^2 \rangle$, 
where $\delta f = f - \langle f \rangle$.
The other type of quantity is r.m.s ``superdeviations'' $\sigma(f)$, 
which measure the super-Hubble content of a field:
\be
\sigma^2(f) = 4\pi \int_{k_{\min}}^{H} dk k^2 P_f(k) \; ,
\label{svar}
\ee
where $P_f$ is the power spectrum of $\delta f = f - \langle f \rangle$,
and $k_{\min} = \pi / L$; $L$ is the radius of the box.

Here we present results for $L = 500\pi$ and $n=8192$.
In Fig. 1 we plot two quantities characterizing the growth of
super-Hubble
perturbations: $\sigma(R)/ 2\kappa^2\langle E \rangle$, 
and $\sigma(\psi) / \langle\psi\rangle$. Both clearly grow, and that growth
begins around the same time as the growth of $\delta\phi$. This is of no
surprise since all these effects are of the same, second order in
$\delta\chi$. (The resonant growth of $\delta\chi$ begins much earlier,
at $t\sim 10$, and ends at $t\sim 80$.)
We also plot the divided by $\langle\psi\rangle$ ordinary r.m.s. deviation 
of $\psi$, which includes all modes. 
We see that at $t > 60$ the full r.m.s. deviation is
barely distinguishable from $\sigma(\psi)$, i.e. at these times almost all of
the variance of $\psi$ is in super-Hubble modes. Same is true for $N$,
but not for example for $R$. This quantifies our conclusion that at times
of interest the metric functions are smooth.

To judge the quality of these results, we plot in Fig. 2 the superdeviations 
of the constraint functions $C_E$ and $C_M$, relative to $\sigma(R)$ and
$\sigma(P)$, respectively. 
The large spikes in $\sigma(C_M)/\sigma(P)$
occur at times when the homogeneous component of $\phi$ reaches a maximum or
minimum, and $\sigma(P)$, by which we divide, becomes especially small.

In Fig. 3 we plot profiles of the lapse function $N(r)$ at
different moments of time.
The values of $t$ give a good approximation to $1/H$, the comoving Hubble radii
at these times (with $H$ defined as above), and one sees that the variation of 
$N$ mostly occurs over super-Hubble scales.
We interpret $N$ crossing zero as formation of a horizon of a black
hole (or, more precisely, what will look as a black hole to an observer
at large $r$): a point where $N=0$, and $\psi$ is
finite, cannot be reached by light from any other point in a finite time.
For a Schwarzschild black hole in asymptotically flat space, the lapse
function, in isotropic coordinates (\ref{met}), crosses zero at $r=r_0/4$,
where $r_0$ is the Schwarzschild radius. In that case, the isotropic coordinates
cover only the region outside the horizon of the black hole, but they cover
it twice. In our calculation, the spacetime is not asymptotically flat, and
at present we are unsure how to interpret the ``inside'' region, where
$N < 0$ (see, however, remarks below).

We find the presence of super-Hubble modes essential: if we reduce the size
of the integration box, so that no such modes are left by the time when
nonlinearity becomes important (by choosing e.g. $L = 40\pi$), $N$ does not
cross zero. For a slightly larger box ($L=70\pi$) we observe formation and 
subsequent disappearance (``evaporation'') of a black hole. These results
correspond well with our observation that $\delta\psi / \psi$ reaches smaller 
values in smaller boxes.

In a large volume, after the zero of $N$ first occurs, it propagates to larger 
$r$, until it becomes comparable to the size of the box, and the constraints 
break down. Shortly before that, however, we observe an interesting change in 
the profiles of
the fields, see Fig. 4. The field $\phi$ in the ``interior'' region becomes
smooth and reminiscent of an excited state 
of a soliton star, cf. ref. \cite{Lee&al},
while the field $\chi$ is pushed out of the ``interior'' (we suppose, by its
repulsive interaction with $\phi$).

In the full three dimensions (as opposed to the spherically symmetric case)
we expect formation of many black holes in different regions of space.
In addition, gravitational waves (absent in spherical symmetry) may be produced in 
these inhomogeneous spacetimes even more efficiently than indicated by
calculations in rigid spacetimes \cite{KTgw}. It remains to see how these various
effects constrain inflationary models in which they occur.

This work was supported 
in part by the U.S. Department of Energy under Grant DE-FG02-91ER40681 (Task B).

\begin{figure}
\leavevmode\epsfysize=3.5in \epsfbox{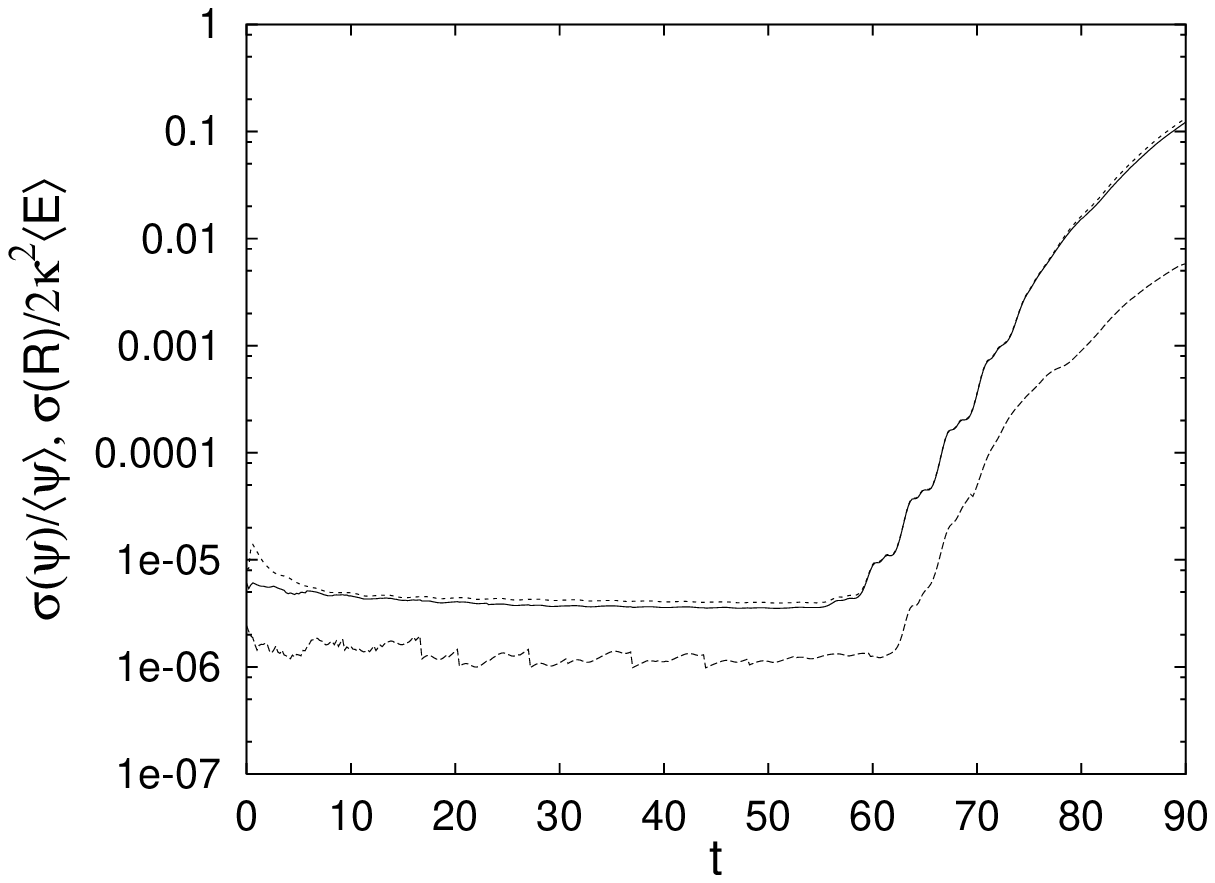}
\vspace*{0.15in}
\caption{
Growth of super-Hubble perturbations: normalized superdeviations 
(see text for the definition) of the metric function
$\psi$ (solid line) and the spatial curvature $R$ (long dashes) as 
functions of the conformal time. The normalized full r.m.s. 
deviation of $\psi$, including all modes, is shown for comparison (short dashes).
}
\label{fig:super}
\end{figure}

\begin{figure}
\leavevmode\epsfysize=3.5in \epsfbox{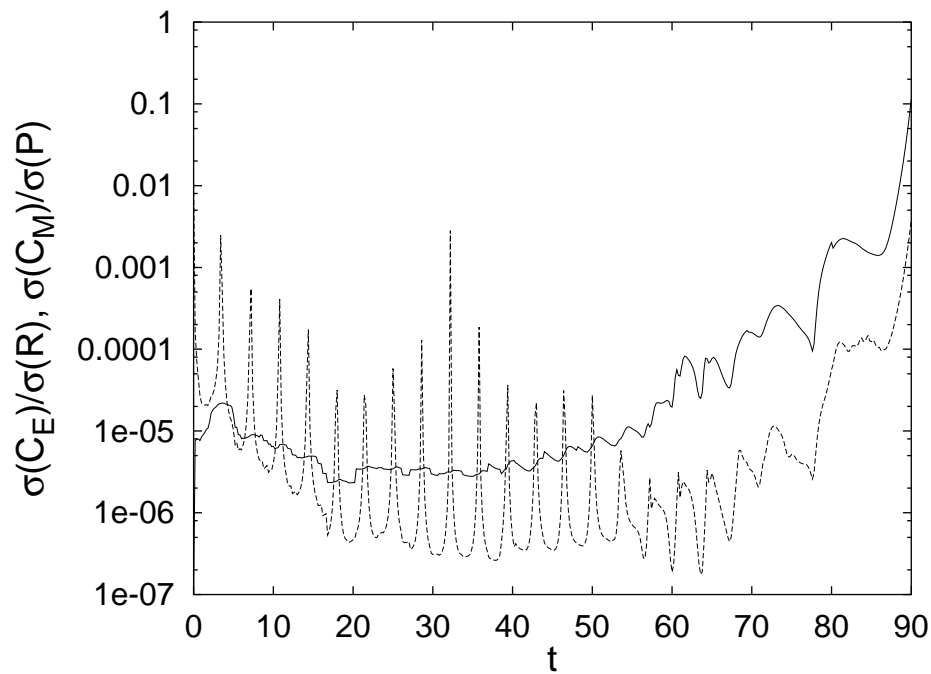}
\vspace*{0.15in}
\caption{
Normalized superdeviations of the constraint functions $C_E$ (solid line) 
and $C_M$ (dashed line) as functions of conformal time.
}
\label{fig:constr}
\end{figure}

\begin{figure}
\leavevmode\epsfysize=3.5in \epsfbox{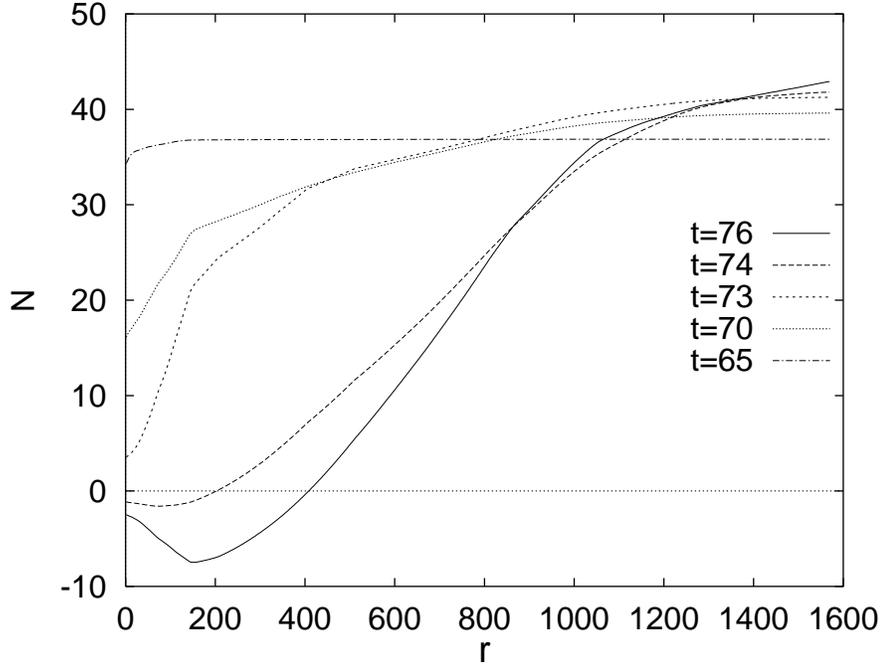}
\vspace*{0.2in}
\caption{
Profiles of the lapse function $N$ at several moments of time. We interpret the
crossing of zero as formation of a black hole.
}
\label{fig:lapse}
\end{figure}

\begin{figure}
\leavevmode\epsfysize=3.5in \epsfbox{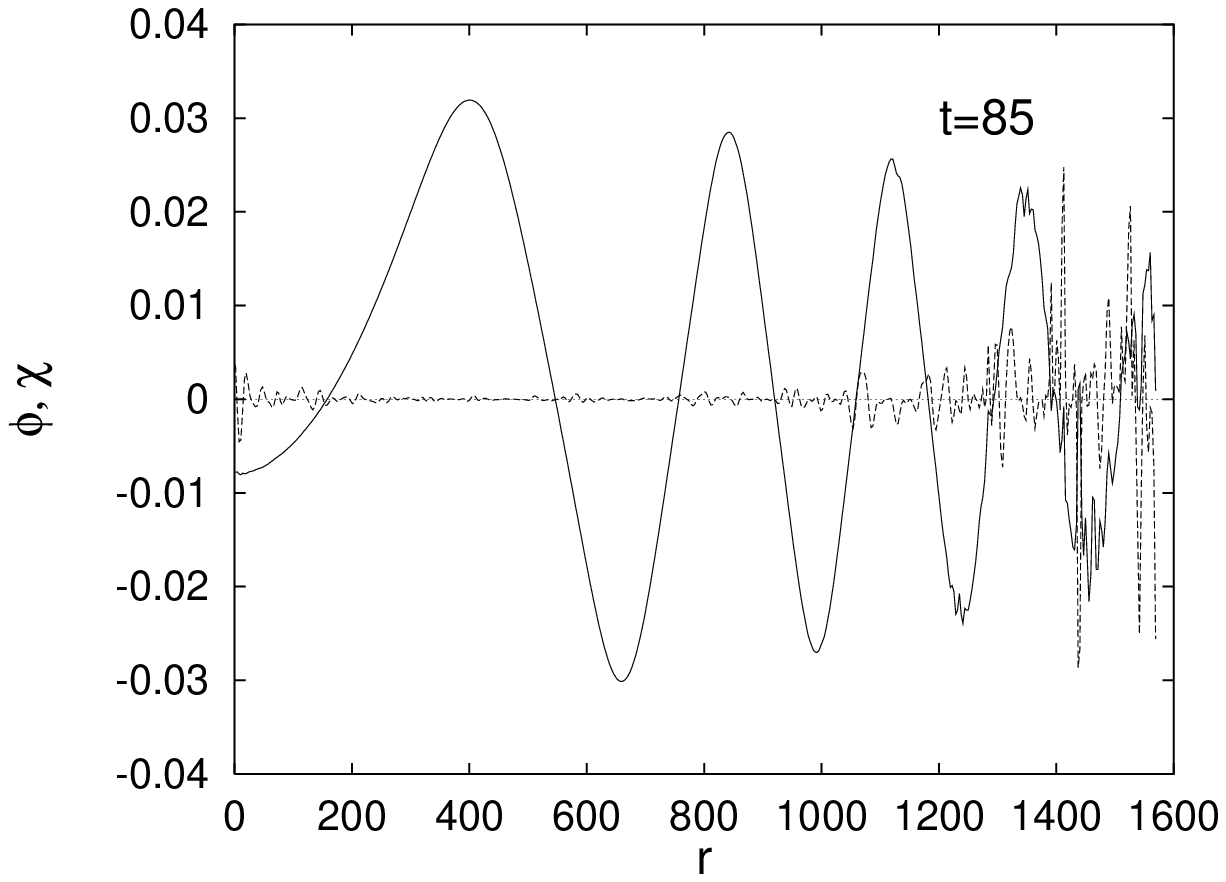}
\vspace*{0.2in}
\caption{
Profiles of the fields $\phi$ (solid line) and $\chi$ (dashed line) at
$t=85$.
}
\label{fig:fields}
\end{figure}


\begin{thebibliography}{10}

\bibitem{books} A. D. Linde, {\em Particle Physics and Inflationary
Cosmology} (Harwood, Chur, Switzerland, 1990); E. W. Kolb and M. S.
Turner, {\em The Early universe} (Addison-Wesley, Redwood City,
California, 1990).

\bibitem{structure} V. F. Mukhanov and G. V. Chibisov, JETP Lett.
{\bf 33}, 532 (1981); S. W. Hawking, Phys. Lett. {\bf 115B}, 295 (1982);
A. A. Starobinsky, Phys. Lett. {\bf 117B}, 175 (1982);
A. H. Guth and S.-Y. Pi, Phys. Rev. Lett. {\bf 49}, 1110 (1982).

\bibitem{param} J. Traschen and R. Brandenberger, Phys. Rev. D
{\bf 42}, 2491 (1990); L. Kofman, A. Linde, and A. Starobinsky,
Phys. Rev. Lett. {\bf 73}, 3195 (1994); Y. Shtanov, J. Traschen, 
and R. Brandenberger, Phys. Rev. D {\bf 51}, 5438 (1995).

\bibitem{KT1} S. Khlebnikov and I. Tkachev, Phys. Rev. Lett. {\bf
77}, 219 (1996).

\bibitem{variety} S. Khlebnikov and I. Tkachev, Phys. Lett. B {\bf
390}, 80 (1997); T. Prokopec and 
T. G. Roos, Phys. Rev. D {\bf 55}, 3768 (1997);
S. Khlebnikov and I. Tkachev, Phys. Rev. Lett. {\bf
79}, 1607 (1997).  

\bibitem{BKM} B. Bassett, D. Kaiser and R. Maartens, Phys. Lett. B
{\bf 455}, 84 (1999).

\bibitem{FB1} F. Finelli and R. Brandenberger, Phys. Rev. Lett. {\bf
82}, 1362 (1999).

\bibitem{BV} B. Bassett and F. Viniegra, Phys. Rev. D 
{\bf 62}, 043507 (2000).

\bibitem{FB2} F. Finelli and R. Brandenberger, hep-ph/0003172, to be
published in Phys. Rev. D.

\bibitem{ZBS} J. P. Zibin, R. Brandenberger, D. Scott, hep-ph/0007219.

\bibitem{GKLS} P. Greene, L. Kofman, A. Linde, and A. Starobinsky, 
Phys. Rev. D {\bf 56}, 6175 (1997).

\bibitem{long} F. Finelli and S. Khlebnikov, in preparation.

\bibitem{EP2} R. Easther and M. Parry, hep-ph/9910441.

\bibitem{GMBT} A. M. Green and K. A. Malik, hep-ph/0008113; 
B. Bassett and S. Tsujikawa, hep-ph/0008328. 

\bibitem{Lee&al} T. D. Lee, Phys. Rev. D {\bf 35}, 3637 (1987); 
R. Friedberg, T. D. Lee, and Y. Pang, Phys. Rev. D {\bf 35}, 3640
(1987) and {\it ibid.}, 3658 (1987).

\bibitem{KTgw} S. Khlebnikov and I. Tkachev, Phys. Rev. D {\bf 56},
653 (1997). 

\end{thebibliography}
\end{document}